PAPER

# Diagnosis of Patients with Viral, Bacterial, and Non-Pneumonia Based on Chest X-Ray Images Using Convolutional Neural Networks

Carlos Arizmendi(✉), Jorge Pinto, Alejandro Arboleda, Hernando González

Universidad Autónoma de Bucaramanga (UNAB), Bucaramanga, Colombia

carizmendi@unab.edu.co

**ABSTRACT**

According to the World Health Organization (WHO), pneumonia is a disease that causes a significant number of deaths each year. In response to this issue, the development of a decision support system for the classification of patients into those without pneumonia and those with viral or bacterial pneumonia is proposed. This is achieved by implementing transfer learning (TL) using pre-trained convolutional neural network (CNN) models on chest x-ray (CXR) images. The system is further enhanced by integrating Relief and Chi-square methods as dimensionality reduction techniques, along with support vector machines (SVM) for classification. The performance of a series of experiments was evaluated to build a model capable of distinguishing between patients without pneumonia and those with viral or bacterial pneumonia. The obtained results include an accuracy of 91.02%, precision of 97.73%, recall of 98.03%, and an F1 Score of 97.88% for discriminating between patients without pneumonia and those with pneumonia. In addition, accuracy of 93.66%, precision of 94.26%, recall of 92.66%, and an F1 Score of 93.45% were achieved for discriminating between patients with viral pneumonia and those with bacterial pneumonia.

**KEYWORDS**

pneumonia, transfer learning (TL), convolutional neural networks (CNN), dimensionality reduction, support vector machine (SVM)

## 1    INTRODUCTION

Pneumonia is a type of acute respiratory infection characterized by inflammation of the distal airway and parenchyma [1, 2]. Inflammation occurs in response to various infectious agents such as viruses, bacteria, and fungi [2]. The severity of pneumonia depends on factors such as preexisting conditions, the site of acquisition, and the infectious agent [2, 3]. Despite the etiological factors of the disease, many cases exhibit similar symptoms. Accurate classification of the type of pneumonia is essential for defining diagnostic and therapeutic actions [4, 5]. Classification of pneumonia depends on the type of causative microorganism and the site of infection acquisition [1, 3]. Bacterial pneumonia is the most common, generally acquired in the community, in healthcare settings, or through aspiration [3]. Viral pneumonia is of special importance due to the resurgence of viruses as causative agents of severe pneumonia, mainly attributed to the introduction of pneumococcal vaccines, which led to an increase in viral infections. This caused the severe acute respiratory syndrome (SARS), with new viruses discovered in the last decade, such as the NL63 and HKU1 coronaviruses, human bocavirus, and COVID-19 [6]. Molecular diagnostic tests have facilitated the epidemiological characterization of these viruses, with chest x-rays confirming pneumonia diagnosis [5, 6]. It's crucial to accurately diagnose the causative agent of pneumonia due to the differences in treatment approaches.

Advancements in artificial intelligence (AI) and deep learning (DL) have enabled the evolution and development of increasingly powerful algorithms that reduce the margin of error in medical image classification tasks [7]. They excel in tasks such as detecting lymph nodes even under unfavorable conditions due to the low contrast of surrounding structures in computed tomography. In [8], the Chest-X-Ray8 database was developed, consisting of 108,948 frontal view x-ray images, with 32,717 having multiple disease labels extracted from radiology report text.

Rajpurkar and colleagues [9] designed a 121-layer convolutional neural network (CNN) named ChexNet, trained on the chest x-ray (CXR) 14 dataset. Yao et al. [10] exploited the interdependencies among target labels when predicting 14 pathological patterns in chest x-rays. This approach yielded improved results in 13 out of 14 outcomes. These advancements highlight the potential of AI and DL to enhance medical image analysis, particularly in tasks that demand high accuracy and nuanced pattern recognition.

Although mortality rates from acute respiratory failure (ARF) have decreased in recent years, the progress hasn't been notably significant since penicillin became a routine treatment [11]. Various international studies show that in developed countries, there will be at least 915,900 cases of respiratory diseases, with pneumonia being the most common, consisting of approximately 83 serotypes differentiated by their capsular polysaccharides. The incidence rate is around six cases per 100,000 inhabitants in the United States [12].

Given the evident challenges posed by ARFs for communities with limited access to specialized studies such as microbiological or x-ray imaging, it's estimated that by 2030, out of 100% of medical care specialists, only 4.3% will be experts in radiology and diagnostic imaging [13]. Consequently, it is essential to develop technological tools that can facilitate the process of analyzing CXR images. This is crucial to address the health issues posed by respiratory diseases in communities with limited access to specialized medical resources. This study implements CNN and transfer learning (TL) using three selected architectures: Resnet-18, Resnet-50, and Incep-Resnet-v2. Dimensionality reduction techniques, Relief and Chi-square [14], were implemented along with support vector machines (SVM) [15] as a complementary classification technique to that performed using CNN. This methodology enables the creation of an AI-based model for diagnosing viral and bacterial pneumonia in patients using x-ray images. The paper is structured into four main sections: introduction, material and methods, results, and discussion.

Recent developments in diagnosing pneumonia from CXR images have been facilitated by the integration of DL and machine learning techniques, as demonstrated by several notable studies. Manickam et al. present a new DL method for detecting pneumonia from CXR images. The aim is to improve accuracy and streamline the diagnosis process. The method uses pre-trained architectures such as ResNet50, InceptionV3, and InceptionResNetV2 to extract spatial and temporal features from the images, leveraging transfer learning. The system uses U-Net architecture-based segmentation to preprocess input images and identify the presence of pneumonia, classifying it as normal or abnormal (bacterial or viral). The study also evaluates the performance of different optimizers, including Adam and stochastic gradient descent (SGD), with varying batch sizes to improve feature extraction and classification accuracy. In a study with convolutional neural networks with TL for pneumonia detection, the InceptionV3 model achieved the best performance with 72.9% accuracy, recall 93.7%, and F1-Score 82%. These findings demonstrate the potential of DL approaches to revolutionize pneumonia diagnosis, offering a promising avenue for improving patient outcomes through more accurate and efficient detection methods [16]. The COVID-19 pandemic has highlighted the need for effective diagnostic tools to combat the virus and its complications, particularly pneumonia. Timely identification of pneumonia in COVID-19 patients is crucial, especially for older adults and those with preexisting medical conditions who are at higher risk. This study utilizes ML techniques and DL treatment in a diagnosis model using TL with a VGG16 net, Adam algorithm serving as the optimizer. The model shows promising performance by accurately predicting pneumonia in COVID-19 patients with an average accuracy of 91.69%, sensitivity of 95.92%, and specificity of 100%. Additionally, the model demonstrates efficient training loss reduction and accuracy enhancement. The study concludes by highlighting the potential of the proposed method as an alternative diagnostic tool for identifying pneumonia cases in COVID-19 patients [17]. Ieracitano et al. design CovNNet, a fuzzy enhanced DL approach for early detection of Covid-19 pneumonia from portable CXR images. The proposed model integrates fuzzy logic with DL to distinguish between CXR images of Covid-19 pneumonia and those with interstitial pneumonias unrelated to Covid-19, addressing the urgent need for efficient diagnosis and triaging of Covid-19 patients. The CovNNet model achieves higher classification performance by combining CXR and fuzzy features within a multilayer perceptron framework. Accuracy rates reach up to 81%, surpassing benchmark DL approaches. The proposed approach's effectiveness is further validated through additional datasets, showcasing its potential for triaging patients in acute settings. Furthermore, the study examines permutation analysis and proposes a straightforward occlusion methodology for explaining decisions. The proposed pipeline can be seamlessly integrated into existing clinical decision support systems. Future research directions include exploring recent performance measures tailored for imbalanced datasets, validating the approach with larger datasets, integrating it with other state-of-the-art AI approaches, and considering legal issues in medical AI to ensure compliance with privacy regulations. Furthermore, this study will investigate hybrid interactions between humans and machine intelligence, as well as interactive machine learning strategies, to improve the process of knowledge discovery and address uncertainties that are inherent in medical datasets [18].

This study highlights the scientific contribution of performing an exhaustive analysis of the execution with different CNNs by reviewing the accuracy, precision, recall, and F1 score to perform the differential diagnosis between viral pneumonia, bacterial pneumonia, and non-pneumonia by implementing feature extraction in the last convolutional layer of each network and classifying it using an SVM, obtaining higher performance than other authors. Additionally, as a scientific novelty in the CNN area, the feature extraction stage was investigated for the variation in the performance of all the computed execution measures by reducing the dimensionality of the extracted features by implementing the Relieff and Chi-Square methodologies; this reduction is the subject of discussions to improve the generalization of the final model and avoid the curse of dimensionality in developments with CNN in general. The objective of the study was to increase most of the performance found in the literature, opening a new study for different dimensionality reduction techniques with higher prestation than statistical methodologies. The final objective of the study is to expand the number of models to be analyzed, obtaining the best diagnostic results, which can be implemented on a cell phone to carry out a pre-diagnosis of support decisions, to help overcome the problem of the lack of specialists in the field, helping people with limited access or resources to early diagnosis of the disease.

### 1.1 Database

In this study, the CXR pneumonia database was utilized, obtained from the collection of images labeled optical coherence tomography (OCT) and CXR images for classification [19]. The database comprises 5864 grayscale CXR images, with labels representing pneumonia: 2580 bacterial, 1500 viral, and 1784 normal cases (see Figure 1). These images are made available for free download on the Melendey Data platform by the same authors.

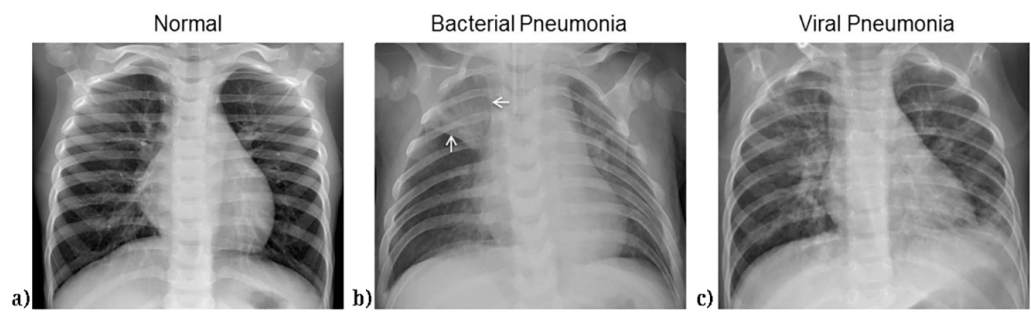

**Fig. 1.** Illustrates the examples of chest x-rays from the dataset [19]. The normal Chest x-ray (a), bacterial pneumonia (b), viral pneumonia (c)

### 1.2 Neural networks and deep learning

Artificial neural networks (ANNs) are a collection of simple elements that operate in parallel to form an artificial computational system. They are designed to mimic the way the human brain learns [20]. ANNs consist of nodes or neurons, which receive a certain number of inputs and apply a function called an activation function to them. The output values determine the activation level of each neuron [21]. Nodes are interconnected with associated weights that interact with the neuron outputs. The number of input neurons corresponds to the system's variables, and the number of neurons in the output layer corresponds to the number of classes. During the learning process, the associated weights are adjusted as the network is trained, enabling ANNs to perform accurately across various problem domains.

**Convolutional neural networks.** The design of CNNs consists of several convolutional layers, pooling layers, and a fully connected layer that forms an ANN to produce the system's output (see Figure 2). The network takes input images and applies convolutional filters to obtain a feature map with a size reduction through pooling layers. This process extracts dominant features and more complex characteristics. Finally, the fully connected output layer comprises a neural network with several neurons equal to the number of classes, computing the system's output [22].

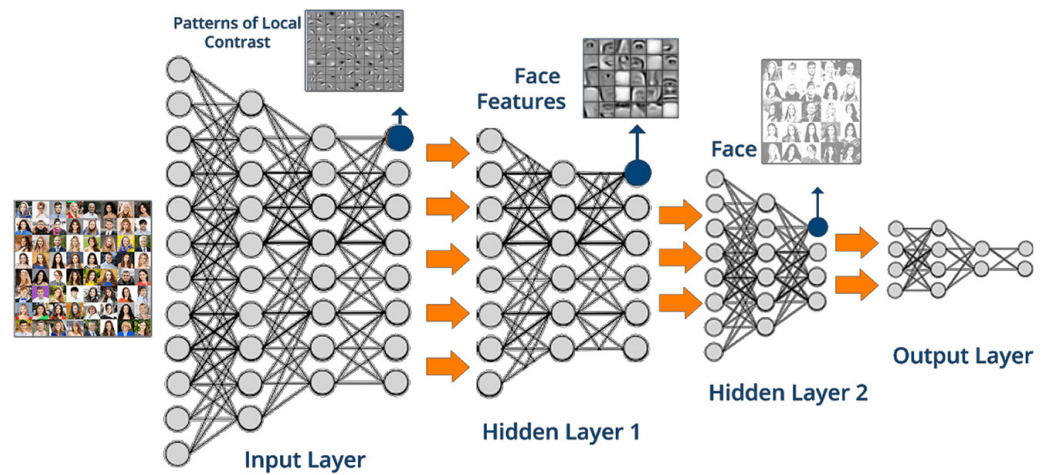

**Fig. 2.** Architecture of a deep convolutional neural network [23]

### 1.3 Support vector machines

Support vector machines map input points, generating a higher dimension through the implementation of kernels. Inside of this higher-dimensional space, an optimal hyperplane is found separating the classes. SVMs learn the decision boundary of the input classes using support vectors, which are patterns acting as critical points in forming a classification decision boundary. Complying with the principle of structural risk minimization (SRM), which aims to minimize a bound on the generalization error of the model instead of minimizing the mean squared error of the training data set. In cases where linear separability isn't possible, non-negative variables are introduced to measure classification errors. These features are mapped to a higher-dimensional space, creating a non-linear classification boundary [24, 25].

Support vector machines were implemented with the radial basis function (RBF) kernel. The hyperparameters of the model were tuned within a Bayesian framework using the best estimated feasible points. This tuning process aimed to minimize the upper confidence bound of the 10-fold cross-validation loss, based on the underlying Gaussian process model of the Bayesian optimization process [26].

### 1.4 Dimensionality reduction

Dimensionality reduction involves transforming high-dimensional data into a lower-dimensional representation retaining meaningful information, ideally with a minimal number of necessary parameters [27]. Feature extraction from deep networks is performed from the last convolutional level of each network before the fully connected classification layer. Containing the most relevant variables for classification by transforming the original images into a more representative feature representation [28]. The extracted features can be directly presented to a classifier or used as a preliminary step before presenting data to the classifier.

In the present study, feature selection methods were implemented, specifically filter type methods: relief and chi-square, along with the elbow method [29], are used to compute the score of each variable. This involved ranking and selecting the highest scoring features while discarding less relevant variables [30, 31]. SVMs were used as the classifier that receives the variables resulting from the dimensionality reduction process.

## 1.5 Methodology

Patterns are balanced using the downsizing method [32, 33], resulting in 1500 patterns in each class, forming a dataset of 4500 images in total. A random 10% of the images are sampled to create a Test 2 set, while the remaining 90% of the data is used for parameter tuning of the models to be implemented. This data is resampled using Hold Out Validation [34], with 60% for the training set, 20% for the validation set, and 20% for the test 1 set. To prevent overfitting, an augmented image datastore is implemented, preprocessing the training data with resizing, rotation, and reflection at each epoch [35].

Transfer learning is implemented [36] to retrain the Resnet-18, Resnet-50, and Inception Resnet V2 networks. Pretrained networks are chosen due to the general ease and speed of adjusting their parameters compared to training a network from scratch with randomly initialized weights. The choice of pre-trained networks is based on their size and performance [37]. The original data images are adjusted to the input sizes of each network: 244 pixels by 244 pixels with three input channels for Resnet-18 and Resnet-50, and 299 pixels by 299 pixels with three channels for Resnet-50. For all networks, the Adam optimizer [38] is used with learning rates of 1e–3, 1e–6, and 1e–9, mini-batches of 32, 16, and eight, validation frequency of 50, validation patience of infinity, sequence padding direction of right, maximum number of epochs of 150, and Gradient Threshold Method of L2 norm [39].

Six experiments (Exp) were conducted as sequential steps to select the network along with the optimal model methodology for classifying all classes. In Exp 1, the results of the three pre-trained networks were compared to clinically distinguish between healthy patients and those with pneumonia. The pneumonia classes were grouped into a metaclass called "Pneumonia" (refer to Table 1). Exp 1 focused on the comparison between the normal class and the pneumonia class.

In Exp 2, features were computed and extracted from the three pre-trained networks from the layer immediately preceding the fully connected layer. These extracted features were fed into the SVM, and the comparison was made between the normal class and the pneumonia class (refer to Table 2). The results of Exp 1 and Exp 2 from the validation set were grouped in Table 3. The Inception Resnet V2 network combined with SVM produced the best results in the validation group. Therefore, Inception Resnet V2 along with SVM was selected for subsequent Exp involving the Normal class and Pneumonia class. Exp 3 involved implementing the Inception Resnet V2 network to compute and extract features from the last convolutional layer before the fully connected layer (i.e., conv_7b). These features were then subjected to dimensionality reduction using the Relieff and Chi-square techniques along with the elbow method. This reduced feature set was used as input for SVM classification. Figure 3 illustrates the feature weights obtained through the Relieff and Chi-square methodologies. Table 4 displays the classification results, showing that Relieff yielded the best results in the validation group but did not surpass the results obtained by the Inception Resnet V2 combined with SVM methodology (refer to Table 4). Due to the clinical importance of diagnosis, Exp 4 to Exp 6 focus on differentiating between the viral pneumonia class and the bacterial pneumonia class to find the optimal model between these two classes. In Exp 4, classification is performed between the viral pneumonia class and the bacterial pneumonia class using the Inception Resnet V2 convolutional network (refer to Table 5). In Exp 5, the Inception Resnet V2 network is used to compute and extract features from the last convolutional layer before the fully connected layer for both the viral pneumonia class and the bacterial pneumonia class. These features are then fed into the SVM (refer to Table 5). Exp 6 involves obtaining computed features from the Inception Resnet V2 network at the conv_7b layer. The dimensionality of these features is reduced using the Relieff technique along with the elbow method (see Figure 4). The resulting reduced features are then presented to the SVM classifier (refer to Table 5).

## 2 RESULTS

The results of the Exp along with the respective classes being compared are presented. For the analysis of the results, the performance metrics, including accuracy, precision, recall, and F1 score, were computed for training validation, test 1, and test 2 sets. Tables 1 and 2 display the results of Exp 1 and Exp 2, respectively.

**Table 1.** Comparative performance analysis between the normal class and pneumonia class for the 3 CNNs (i.e., Exp 1)

| | | Performance | | | |
|---|---|---|---|---|---|
| | | Accuracy | Precision | Recall | F1 Score |
| Resnet-18 | Training | 99.90% | 99.90% | 99.90% | 99.90% |
| | Validation | 82.86% | 82.86% | 82.86% | 82.86% |
| | Test 1 | 88.20% | 88.20% | 88.20% | 88.20% |
| | Test 2 | 84.44% | 84.44% | 84.44% | 84.44% |
| Resnet-50 | Training | 99.95% | 99.95% | 99.95% | 99.95% |
| | Validation | 85.30% | 85.30% | 85.30% | 85.30% |
| | Test 1 | 88.60% | 88.60% | 88.60% | 88.60% |
| | Test 2 | 85.78% | 85.78% | 85.78% | 85.78% |
| Inception Resnet V2 | Training | 99.95% | 99.95% | 99.95% | 99.95% |
| | Validation | 89.56% | 89.56% | 89.56% | 89.56% |
| | Test 1 | 91.11% | 91.11% | 91.11% | 91.11% |
| | Test 2 | 92.00% | 92.00% | 92.00% | 92.00% |

**Table 2.** Analysis between the normal class and pneumonia class using the SVM classifier with extracted features from the 3 CNNs (i.e., Exp 2)

| | | Performance | | | |
|---|---|---|---|---|---|
| | | Accuracy | Precision | Recall | F1 Score |
| Resnet-18 | Training | 100% | 100% | 100% | 100% |
| | Validation | 85.22% | 85.22% | 85.22% | 85.22% |
| | Test 1 | 87.56% | 87.56% | 87.56% | 87.56% |
| | Test 2 | 87.56% | 87.56% | 87.56% | 87.56% |
| Resnet-50 | Training | 100% | 100% | 100% | 100% |
| | Validation | 89.00% | 89.00% | 89.00% | 89.00% |
| | Test 1 | 86.90% | 86.90% | 86.90% | 86.90% |
| | Test 2 | 88.40% | 88.40% | 88.40% | 88.40% |
| Inception Resnet V2 | Training | 100% | 100% | 100% | 100% |
| | Validation | 91.02% | 91.02% | 91.02% | 91.02% |
| | Test 1 | 91.20% | 91.20% | 91.20% | 91.20% |
| | Test 2 | 92.30% | 92.30% | 92.30% | 92.30% |

Table 3. Grouped validation results from Exp 1 and Exp 2

| Performance | Resnet-18 | | Resnet-50 | | Inception Resnet V2 | |
|---|---|---|---|---|---|---|
| | Exp 1 | Exp 2 | Exp 1 | Exp 2 | Exp 1 | Exp 2 |
| Accuracy | 82.86% | 85.22% | 85.3% | 89.00% | 89.56% | 91.02% |
| Precision | 93.11% | 92.35% | 95.75% | 94.11% | 96.33% | 97.73% |
| Recall | 94.33% | 97.29% | 94.66% | 95.42% | 96.33% | 98.03% |
| F1 Score | 93.72% | 94.76% | 95.20% | 94.76% | 96.33% | 97.88% |

Figure 3 illustrates the outcomes of Exp 3. Displaying the variables sorted from highest to lowest scores for both the Relieff and Chi-square methods. The elbow method was applied for dimension reduction, condensing the initial 100,000 variables to 6,000 and 15,000 variables, respectively. These reduced variables were then presented to the SVM. Table 4 outlines the results of combining the Inception Resnet V2 network with the SVM, implementing the reduced variables.

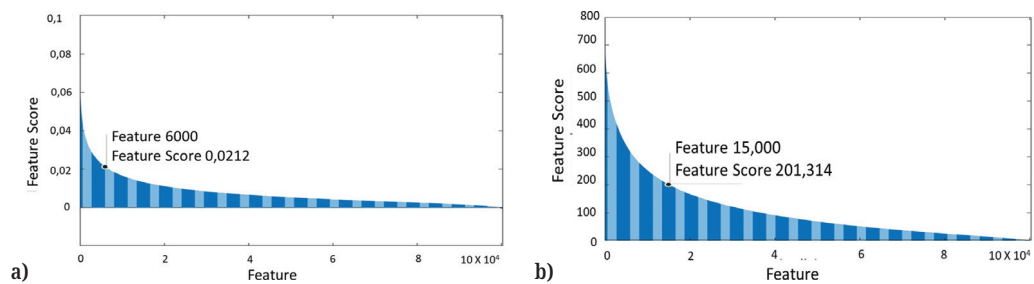

Fig. 3. Ranked activation variable scores implementing the methodologies a) Relief and b) Chi-square

Table 4. Performance of the reduced dataset applying Relief and Chi-square, evaluated with SVM

| | | Performance | | | |
|---|---|---|---|---|---|
| | | Accuracy | Precision | Recall | F1 Score |
| Inception Resnet V2 Relief | Training | 100% | 100% | 100% | 100% |
| | Validation | 90.53% | 90.53% | 90.53% | 90.53% |
| | Test 1 | 91.22% | 91.22% | 91.22% | 91.22% |
| | Test 2 | 90.22% | 90.22% | 90.22% | 90.22% |
| Inception Resnet V2 Chi Square | Training | 100.00% | 100.00% | 100.00% | 100.00% |
| | Validation | 90.26% | 90.26% | 90.26% | 90.26% |
| | Test 1 | 90.56% | 90.56% | 90.56% | 90.56% |
| | Test 2 | 90.00% | 90.00% | 90.00% | 90.00% |

The validation set results from Table 4 illustrate the combination of the Inception Resnet V2 network, dimensionality reduction using Relief, and SVM yielding the highest performance in most of the execution metrics. In Exp 6, the analysis of dimensionality reduction was performed on features extracted from the Inception Resnet V2 network, applying the elbow method. This resulted in a total of 6,000 relevant variables out of the initial 25,000 variables for both viral pneumonia and bacterial pneumonia classes (see Figure 4). These variables were then presented to the SVM classifier. The outcomes of Exp 4, Exp 5, and Exp 6 are summarized in Table 5.

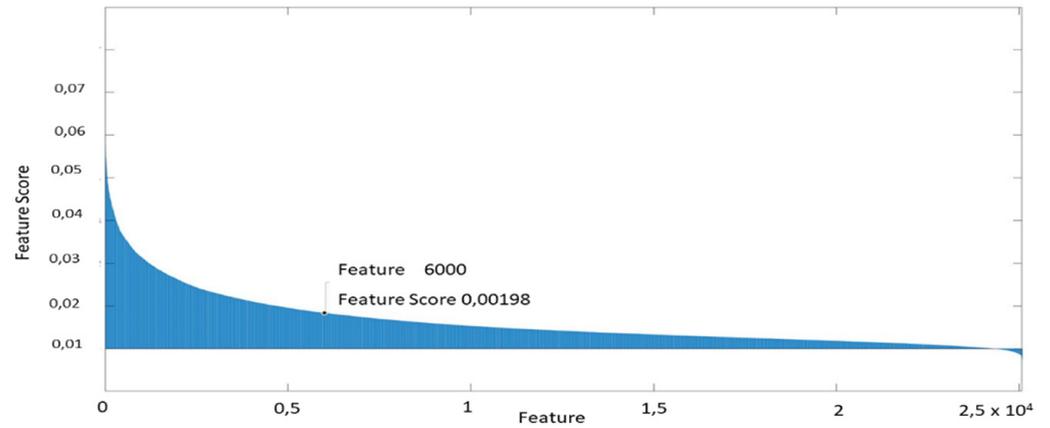

**Fig. 4.** Ranked activation variable scores using the Relief methodology from Exp 6

**Table 5.** Performance of three experiments between viral pneumonia class and bacterial pneumonia class

|  |  | Performance | | | |
|---|---|---|---|---|---|
|  |  | **Accuracy** | **Precision** | **Recall** | **F1 Score** |
| Resnet-18 | Training | 98.95% | 98.95% | 98.95% | 98.95% |
|  | Validation | 92.95% | 92.95% | 92.95% | 92.95% |
|  | Test 1 | 91.25% | 91.25% | 91.25% | 91.25% |
|  | Test 2 | 92.17% | 92.17% | 92.17% | 92.17% |
| Resnet-50 | Training | 100.00% | 100.00% | 100.00% | 100.00% |
|  | Validation | 93.66% | 93.66% | 93.66% | 93.66% |
|  | Test 1 | 94.25% | 94.25% | 94.25% | 94.25% |
|  | Test 2 | 93.50% | 93.50% | 93.50% | 93.50% |
| Inception Resnet V2 | Training | 100.00% | 100.00% | 100.00% | 100.00% |
|  | Validation | 92.77% | 92.77% | 92.77% | 92.77% |
|  | Test 1 | 90.33% | 90.33% | 90.33% | 90.33% |
|  | Test 2 | 92.17% | 92.17% | 92.17% | 92.17% |

Table 5 encloses the best validation set results from the six conducted Exp. It is observed that the results in Table 5, Exp 2, and Exp 5 achieve the top performance for classification between the normal class and the pneumonia class, viral pneumonia class, and bacterial pneumonia class, respectively.

## 3    DISCUSSIONS

The study illustrates the Inception Resnet-V2 network achieving the best results in the analysis of the three proposed networks in Exp 1. Therefore, it is recommended to conduct further research using different types of networks with superior characteristics to analyze whether they can enhance the results obtained in the Exp conducted so far.

In Exp 2, when extracting parameters from the last convolutional layer before the fully connected layer and presenting to the SVM, an improvement in classification results is observed compared to the results obtained implementing CNN without feature extraction. This enhances the generalization of the system. It is recommended to conduct a study using different SVM kernels and various classification systems to compare their performance in terms of execution metrics and generalization.

By performing dimensionality reduction, the number of variables computed for the SVM classifier is reduced by 94% and 75%, respectively, when using the Relief and Chi-square methods in the case of distinguishing between the normal class and the pneumonia class. In this study, the Relief technique achieves better classification results than the Chi-square technique, although it does not manage to surpass the results of the CNN combined with SVM from Exp 2.

In general, dimensionality reduction methods decreased the number of variables presented to the classifier, thereby enhancing system generalization and avoiding the curse of dimensionality. In this study, the Relief and Chi-square techniques did not manage to improve the performance achieved by the Exp conducted without dimensionality reduction. It's advisable to conduct a study using different techniques, possibly wrapper type techniques, which tend to select variables that enhance system performance. These techniques may have more precise rules for selecting the number of relevant variables than the ones implemented.

After conducting multiple Exp, the training, validation, and Test 1 sets tend to exhibit some overlapping input patterns across them. Therefore, the Test 2 set was used for evaluation, resulting in total average variation of all the performance metrics between Tests 1 and 2 of 3.74%, 4.28%, and 2.38% for the results in Tables 1 and 2 for the Resnet-18, Resnet-50, and Inception Resnet V2 networks, respectively. In Table 3, an average variation of 3.51% and 1.00% is observed for the tests performed using Relief and Chi-square techniques, respectively. Table 5 shows average variation of 1.00%, 0.95%, and 2.11% for Exp 4, Exp 5, and Exp 6, respectively.

The results show the need to expand the number of models to be analyzed and the number of dimensionality reduction techniques to obtain better results with lower computational cost in order to be implemented on a cell phone to perform a diagnosis that helps to overcome the problem of lack of specialists in the field and assist people with limited access or resources to early diagnosis of the disease.

## 4  CONCLUSION

In this study, a multiclass CXR analysis was conducted using DL techniques, employing CNNs. Various methodologies, including TL, feature extraction, data augmentation, and dimension reduction by applying filter-like methods, were employed. Evaluation metrics such as accuracy, recall, precision, and F1 score were used to select models with the best performance for classifying the normal class, viral pneumonia class, and bacterial pneumonia class.

Optimal results were observed with the Inception Resnet-V2 network, exhibiting the best performance metrics in the validation set. This improvement in generalization error was significant as it reduced the disparity between Training and Test 2 results in the same Exp.

The results obtained in the study demonstrate the optimal performance of CNNs when their features are extracted and presented to the SVM classifier.

Exp 3 emerged as the best model for classifying between the normal class and the pneumonia class, achieving an accuracy of 91.02% and an F1 score of 97.88%. In Exp 5, the best model for classifying between viral pneumonia class and bacterial pneumonia class was obtained, achieving an accuracy of 92.77% and an F1 score of 92.66%.

## 5  ACKNOWLEDGMENTS

We thank Goldbaum, Michael; Kermany, Daniel; and Zhang, Kang, for the work done in collecting the database, labeling, and preprocessing it.

## 7 AUTHORS

**Carlos Arizmendi** received B.Eng, degree in electronic engineering from Universidad Industrial de Santander (UIS), Bucaramanga, Colombia in 1997; in 2008 received the diploma of advanced studies in the doctorate of biomedical engineering at Universidad Politécnica de Catalunya (UPC), Barcelona, Spain; in 2012 received the PhD in artificial intelligent from UPC. He is currently titular professor of the mechatronics engineering program and biomedical engineering program at Universidad Autónoma de Bucaramanga (UNAB), also is director of the Research Group in Control and Mechatronics (GICYM) at UNAB, creator of the biomedical engineering program at UNAB. His research interests include machine learning, deep learning, signal treatments, electronics, mechatronics and biomedical devices (E-mail: carizmendi@unab.edu.co).

**Jorge Pinto** received the B.Eng., degree in biomedical engineering from UNAB, studying the specialization in Management of Health Institutions at the University of Santander UDES with studies in quality management systems under the ISO9001 standard; trained and experienced in coordinating and performing maintenance on low and high complexity biomedical equipment (E-mail: jpinto520@unab.edu.co).

**Alejandro Arboleda** is the Director of the Biomedical Engineering Program at the Autonomous University of Bucaramanga, doctoral student in biomedical engineering at the UPC, Barcelona-Spain. His research interests include the analysis and treatment of biomedical signals and artificial intelligence (E-mail: aarboleda@unab.edu.co).

**Hernando González** received B.Eng. degree in electronic engineering from UIS, master in electronic engineering from UIS. He is a student of doctorate in engineering from UNAB and is currently working as an Associate Professor of the mechatronics engineering program at UNAB and member of GICYM, His research interests include machine learning, control systems, signal treatments, robotics, electronics and mechatronics devices (E-mail: hgonzalez7@unab.edu.co).